\documentclass[%
 aip,
 amsmath,amssymb,
 preprint,%
author-numerical,%
]{revtex4-1}

\usepackage{graphicx}% Include figure files
\usepackage{dcolumn}% Align table columns on decimal point
\usepackage{bm}% bold math
\usepackage{color}
\usepackage[utf8]{inputenc}
\usepackage[T1]{fontenc}
\usepackage{mathptmx}

\newcommand{\BS}[1]{{\color{black}#1}}
\graphicspath{{./}}

\begin{document}

%\preprint{AIP/123-QED}

\title{Stability of slip channel flow revisited}
% Force line breaks with \\

\author{Chunshuo Chai}
\author{Baofang Song}%
 \email{baofang_song@tju.edu.cn}
\affiliation{ 
Center for Applied Mathematics, Tianjin University, Tianjin 300072, China%\\This line break forced with \textbackslash\textbackslash
}%

\begin{abstract}
In this work, we revisit the temporal stability of slip channel flow. Lauga \& Cossu (Phys. Fluids $\bm 17$, 088106 (2005)) and Min \& Kim (Phys. Fluids $\bm 17$, 108106 (2005)) have investigated both modal stability and non-normality of slip channel flow and concluded that the velocity slip greatly suppresses linear instability and only modestly affects the non-normality. Here we study the stability of channel flow with streamwise and spanwise slip separately as two limiting cases of anisotropic slip and explore a broader range of slip length than previous studies did. We find \BS{that, with sufficiently large slip,} both streamwise and spanwise slip trigger three-dimensional leading instabilities. Overall, the critical Reynolds number is only slightly increased by streamwise slip, whereas it can be greatly decreased by spanwise slip. Streamwise slip suppresses the non-modal transient growth, whereas spanwise slip enlarges the non-modal growth although it does not affect the base flow. Interestingly, as the spanwise slip length increases, the optimal perturbations exhibit flow structures different from the well-known streamwise rolls. \BS{However, in the presence of equal slip in both directions, the three-dimensional leading instabilities disappear and the flow is greatly stabilized.} The results suggest that earlier instability and larger transient growth can be triggered by introducing anisotropy in the velocity slip.
\end{abstract}

\maketitle

\section{Introduction}{\label{sec:intro}}
Fully developed channel flow becomes linearly unstable above $Re\simeq 5772$, but a subcritical transition to turbulence can occur way below this Reynolds number at about $Re$=660 \cite{Tao2018}. The non-normality of the linearized governing equation, via which small disturbances can be transiently amplified by a large factor, explains a possible energy growth mechanism in the subcritical transition to turbulence \cite{Trefethen1993, Reddy1993, Schmid1994, Schmid2007}.

The linear stability and non-normality of channel flow with no-slip boundary condition have been well documented. However, velocity slip of viscous flow can occur on super-hydrophobic surfaces, such as lotus leaves or some specially textured surfaces that can trap air in the micro- and nano-structures on the surfaces and cause velocity slip at the liquid-air interfaces. This velocity slip is usually characterized by a parameter called effective slip length. Though generally very small on normal surfaces, effective slip lengths as large as hundreds of micron have been achieved in exerpiments \cite{Lee2008, Lee2009}. The reader is referred to \cite{Voronov2008, Chattopadhyay2018} and the references therein for a more comprehensive discussion on the achieved slip lengths in experiments. This large slip length renders boundary velocity slip relevant at least to low Reynolds number flows in small systems. There are many numerical and theoretical works on modeling the velocity slip on super-hydrophobic surfaces and on the effects of velocity slip on fluid transport in laminar and fully turbulent flows \cite{Lauga2003, Bazant2008, kamrin2011,Park2013, Seo2016, Yu2016, Kumar2016, Aghdam2016}. However, for stability analysis, usually, a simplification of the complex boundary condition is adopted which treats the complexity resulting from the texture structures and their interaction with flows as an effective homogeneous slip length. Though could be questionable for turbulent flows \cite{Seo2016}, a homogeneous effective slip length in combination with the Navier slip boundary condition has been shown to apply to various flow problems \cite{Gersting1974, Vinogradova1999, Voronov2008, Seo2016}. Based on this simplified boundary condition, linear stability analysis of many flows has been carried out \cite{Chu2004, Lauga2005, Min2005, Ghosh2014a, Ghosh2014b, Chattopadhyay2018, Ghosh2015, Chattopadhyay2017, Chakraborty2019}. Among these studies, some were dedicated to investigations of linear instability of single phase channel flow \cite{Gersting1974, Lauga2005, Chu2004, Min2005,Sahu2008}. These authors concluded that velocity slip suppresses linear instability. However, they only investigated the stability of two-dimensional (2-D) modes with zero spanwise velocity, which are known to be the leading unstable modes in the no-slip case. Lauga \& Cossu (2005) \cite{Lauga2005} and Min \& Kim (2005) \cite{Min2005} also studied the non-modal transient growth of slip channel flow. They showed that streamwise velocity slip suppresses the transient growth, whereas Min \& Kim (2005) \cite{Min2005} found that spanwise slip has the opposite effect. Nevertheless, they all concluded that both streamwise and spanwise slip do not affect the flow structure of the most amplified perturbations which are streamwise rolls as in the no-slip case. Besides linear analysis, Min \& Kim (2005) \cite{Min2005} also carried out direct numerical simulations and studied the effects of velocity slip on the transition to turbulence. They showed that an earlier transition was triggered by spanwise slip while streamwise slip delays transition. 

We follow Min \& Kim (2005) \cite{Min2005} and consider the slip of streamwise and spanwise velocity components separately as the limiting cases of anisotropic slip at the channel wall. \BS{With increasingly large slip length achieved in experiments, the effect of velocity slip becomes important even in flow problems beyond micro-fluidics. For example, in channel flow with a gap width on the order of millimeter, slip lengths of tens to hundreds of micron \cite{Lee2008,Lee2009} can reach as large as tenths if normalized by the gap width, much larger than the previously investigated \cite{Lauga2005,Min2005, Ghosh2014b}.} Therefore, in this work, we perform studies in a broader slip length range the effects of the anisotropy in velocity slip on the linear stability and non-modal transient growth. We will show that, as the slip length increases, both streamwise and spanwise slip can trigger different types of linear instability and different optimal non-modal perturbations compared to previous studies.

\section{Methods}\label{sec:methods}

\begin{figure}[ht]
\centering
\includegraphics[width=0.7\linewidth]{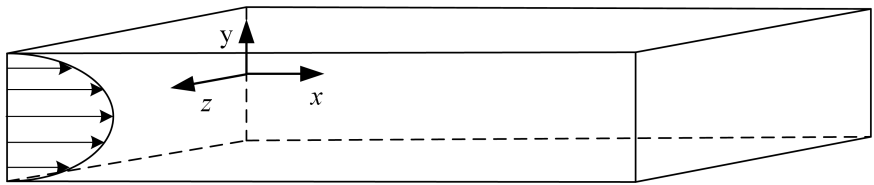}
\caption[transient growth Re=3000 alpha=1 beta=0]{
    \label{fig:geometry} 
     \BS{The geometry and axes system for the channel flow considered in this work. Two infinite plates are separated by a distance of $2h$ and the flow is driven between the plates in the positive $x$ direction.}
} 
\end{figure}

We consider the nondimensional incompressible Navier-Stokes equations
\begin{equation}\label{equ:NS}
 \frac{\partial \bm u}{\partial t}+{\bm u}\cdot\bm{\nabla}
{\bm u}=-{\bm{\nabla}p}+\frac{1}{Re}{\bm\nabla^2}{\bm u}, \;
\bm{\nabla}\cdot{\bm u}=0
\end{equation}
for channel flow in Cartisian coordinates $(x, y, z)$, where $\bm u$ denotes velocity, $p$ denotes pressure and $x$, $y$ and $z$ denote the streamwise, wall-normal and spanwise coordinates, respectively. Velocities are normalized by $U=3U_b/2$ where $U_b$ is the bulk speed, length by half gap width $h$ and time by $h/U$. The Reynolds number is defined as $Re=Uh/\nu$ where $\nu$ is the kinematic viscosity of the fluid. The origin of the y-axis is placed at the channel center (see FIG. \ref{fig:geometry}). 

We use the Navier slip boundary condition at the channel wall for streamwise and spanwise velocities
\begin{equation}\label{equ:BC}
\left(\lambda_{\{x,z\}}\frac{\partial{u_{\{x,z\}}}}{\partial n}+u_{\{x,z\}}\right)|_{y=\pm 1}=0 
\end{equation}
where $n$ is the outward wall-normal direction and $\lambda_x$ and $\lambda_z$ are streamwise and spanwise slip lengths, respectively. In fact, the slip lengths need not to be equal on top and bottom walls \cite{Lauga2005}, however, we only consider the case with equal slip length for both walls in the present work. Impermeability boundary condition is imposed for the wall-normal velocity component, i.e., $u_y(x,\pm 1,z,t)=0$.  

\subsection{The linearization}\label{sec:linearization}

We denote the fully developed base flow as ${\bm U}_b=U_b(y){\bm e}_x$, where ${\bm e}_x$ is the unit vector in the streamwise direction. Introducing small disturbances $\bm u'$ and linearizing the Navier-Stokes equations about the base flow, we obtain the governing equations for $\bm u'$ as the following,

\begin{equation}\label{equ:LNS}
 \frac{\partial \bm u'}{\partial t}+{\bm u'}\cdot\bm{\nabla}{{\bm U}_b} 
+{\bm U}_b\cdot\bm{\nabla}{\bm u'} =-{\bm{\nabla}p'}+\frac{1}{Re}{\bm\nabla}^2{\bm u'}, \;
\bm{\nabla}\cdot{\bm u'}=0.
\end{equation}
The boundary condition (\ref{equ:BC}) is imposed for $\bm u'$. In the following, we will drop the superscript $'$ for all the perturbative quantities.

\subsection{The adjoint system}
Here we choose to adopt the adjoint-based method described in \cite{Barkley2008} for our non-modal analysis, which can easily work with primitive variables, unsteady base flow, complex boundary condition and geometry and therefore is more versatile. \BS{Besides, this method can obtain the optimal perturbation as well as its time evolution simultaneously.} 

Following Barkley et al. (2008) \cite{Barkley2008}, the adjoint system of Eqs. (\ref{equ:LNS}) can be derived as
\begin{equation}\label{equ:LNS_adjoint}
 -\frac{\partial \bm u^*}{\partial t}+{\bm u^*}\cdot(\bm{\nabla}{{\bm U}_b})^T 
-{\bm U}_b\cdot\bm{\nabla}{\bm u^*} =-{\bm{\nabla}p^*}+\frac{1}{Re}{\bm\nabla^2}{\bm u^*}, \;
\bm{\nabla}\cdot{\bm u^*}=0
\end{equation}
with the same boundary condition (\ref{equ:BC}) for $\bm u^*$ at the channel wall, \BS{where the starred quantities are the respective adjoint of those in Eqs. (\ref{equ:LNS}) and the superscript $T$ denotes matrix transpose.} 

\subsection{Optimal energy growth}
Denoting the kinetic energy of $\bm u(\tau)$ at time $\tau$ as \BS{$E(\tau)=\left\|\bm u(\tau)\right\|_2=\int_V\bm u(\tau) \cdot \bm u(\tau) dV$}, where $V$ denotes the integration volume, the maximum possible energy growth at time $\tau$ of an initial perturbation $\bm u(0)$  
\begin{equation}\label{equ:TG_definition}
G(\tau)=\underset{{\left\|\bm u(0)\right\|_2\neq 0}}{\text{max}}\frac{E(\tau)}{E(0)}
\end{equation} 
can be calculated as the maximum eigenvalue of the operator $A^*(\tau)A(\tau)$, where $A(\tau)$ and $A(\tau)^*$ are the action operators to map $\bm u(0)$ to $\bm u(\tau)$ according to Eqs. (\ref{equ:LNS}) and $\bm u^*(0)$ to $\bm u^*(\tau)$ according to Eqs. (\ref{equ:LNS_adjoint}), respectively \cite{Barkley2008}. 

This method does not explicitly derive $A(\tau)$ and $A^*(\tau)$, instead, directly evaluates the output of the action $A^*(\tau)A(\tau)$ given an input $\bm u(0)$ by time-stepping Eqs. (\ref{equ:LNS}) forward from $t=0$ to $t=\tau$ and Eqs. (\ref{equ:LNS_adjoint}) backward from $t=\tau$ to $t=0$. Subsequently, the Krylov subspace method is used to iteratively approximate the maximum eigenvalue of $A^*(\tau)A(\tau)$. In this way, the boundary geometry, incompressibility constraint and boundary condition are taken care of numerically by the solvers for the Navier-Stokes and adjoint equations. \BS{Surely it is more computationally expensive than the usual algorithm based on singular value analysis of the linearized Navier-Stokes operator \cite{Reddy1993,Lauga2005}, but still affordable for the current problem at relatively low Reynolds numbers.}

\subsection{Discretization and time-stepper}
The linearized incompressible systems (\ref{equ:LNS}) and (\ref{equ:LNS_adjoint}) are solved using a Fourier spectral-Chebyshev collocation method. In the streamwise and spanwise directions, periodic boundary conditions are imposed and Fourier spectral method is used for the spatial discretization. In the wall normal direction, Chebyshev-Gauss-Lobatto grid points and Chebyshev-collocation method \cite{Trefethen2000} are used for the spatial discretization. For the channel geometry, the ($\alpha, \beta)$ mode of velocity and pressure is expressed as
\begin{equation}
B(x,y,z,t)_{(\alpha, \beta)}={\hat B}_{(\alpha,\beta)}(y,t)e^{(i\alpha x+i\beta z)} + cc.,
\end{equation}
where $\alpha$ and $\beta$ are the streamwise and spanwise wave numbers, respectively, ${\hat B}_{(\alpha,\beta)}$ is the Fourier coefficient of the mode $(\alpha,\beta)$ and $cc.$ represents complex conjugate. The integration in time is performed using a second-order-accurate Adams-Bashforth/backward-differentiation scheme and the incompressibility condition is imposed using the projection method proposed by \cite{Hugues1998}.

\subsection{Velocity-vorticity formulation}
For the study of linear instability of the flow, we need to search for unstable eigenvalues with four varying parameters, i.e., $Re$, $\alpha$, $\beta$ and slip length. The vast parameter space to explore makes the adjoint method described above expensive, especially when the modal growth is itself very slow near the critical Reynolds numbers and when $Re$ is high such that it takes very long time for the modal growth to outweigh the non-modal one. Therefore, we adopt the velocity-vorticity formulation of the linearized Navier-Stokes equations \cite{Reddy1993} and directly calculate the eigenvalues of the linear operator. 
\BS{The linearized equations in this formulation read
\begin{eqnarray}
 \left(\frac{\partial}{\partial t}+U_b\frac{\partial}{\partial x}\right)\nabla^2u_y - \frac{d^2U_b}{dy^2}\frac{\partial u_y}{\partial x} & =\frac{1}{Re}\nabla^4u_y, \\
 \left(\frac{\partial}{\partial t}+U_b\frac{\partial}{\partial x}\right)\eta + \frac{dU_b}{dy}\frac{\partial u_y}{\partial z} &=\frac{1}{Re}\nabla^2\eta,
\end{eqnarray}
where $\eta=\partial u_x/\partial z-\partial u_z/\partial x$ is the $y$-component of the vorticity. 
Using the incompressibility condition, $u_x$ and $u_z$ can be derived in spectral space as
\begin{eqnarray}
\hat u_x &=\frac{1}{i(\alpha^2+\beta^2)}\left(\beta \hat\eta -\alpha\frac{\partial\hat u_y}{\partial y}\right)\\
\hat u_z &=\frac{1}{i(\alpha^2+\beta^2)}\left(-\alpha\hat\eta-\beta \frac{\partial\hat u_y}{\partial y}\right).
\end{eqnarray}
Further, the boundary condition for $\eta$ can be derived using the slip boundary condition (\ref{equ:BC}). It should be noted that $\eta$ and $u_y$ are coupled via this boundary condition. Therefore, we have four boundary conditions coupling $\eta$ and $u_y$, two on each wall, and $u_y(y=\pm 1)=0$, which together are the six boundary conditions needed for our system. The same Fourier spectral-Chebyshev collocation discretization for the adjoint method is used here for discretizing the 	 linear operator.}

\section{Results}
\subsection{Method validation}
\begin{figure}
\centering
\includegraphics[width=0.9\linewidth]{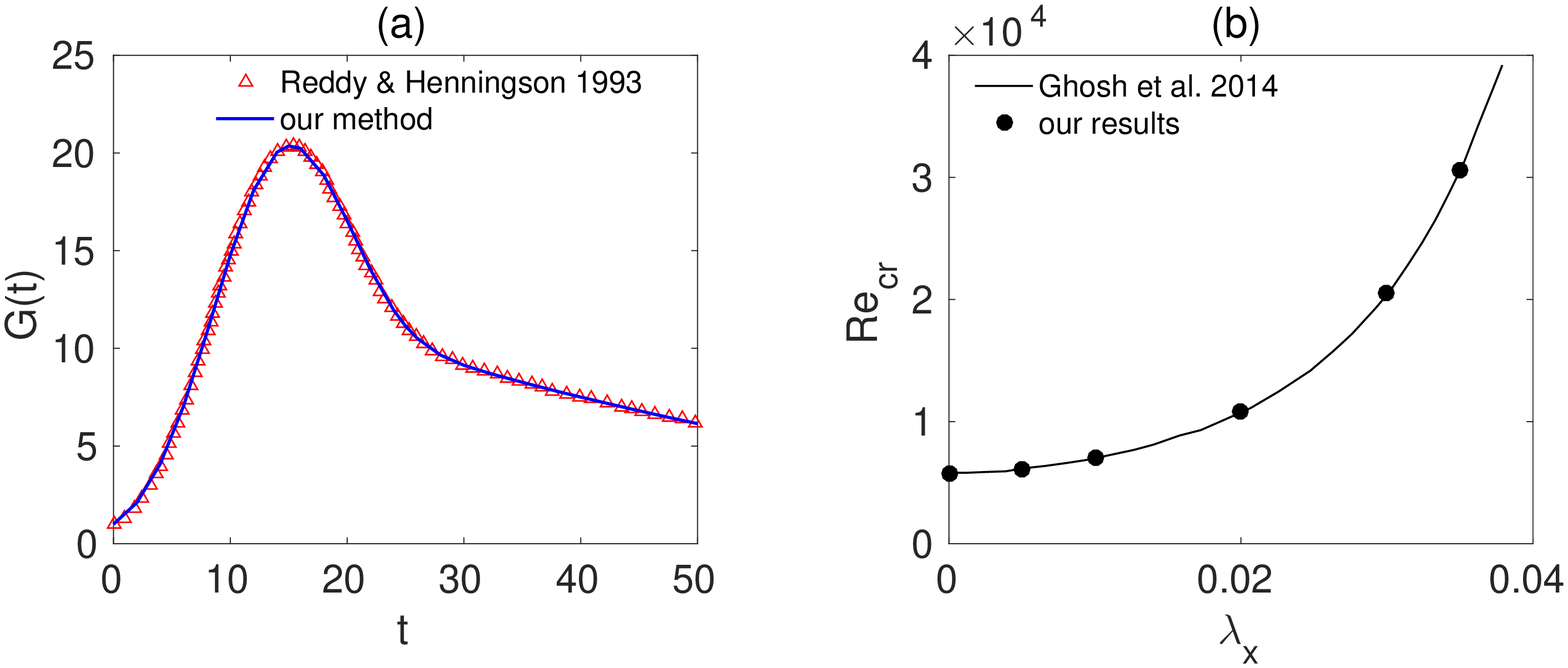}
\caption[transient growth Re=3000 alpha=1 beta=0]{
    \label{fig:validation} 
     (a) $G(t)$ for Re=3000 with $\alpha=1$ and $\beta=0$. Our result (line) is compared with the result of Reddy \& Henninson (1993) \cite{Reddy1993} (symbols). \BS{(b) The critical Reynolds number of 2-D ($\beta=0$) modes as a function of $\lambda_x$. The reference values from Ghosh {\it et al.} 2014 \cite{Ghosh2014b} are plotted as a solid line.} 
} 
\end{figure}

We validated our adjoint method against the transient growth calculated by \cite{Reddy1993} for channel flow with no-slip boundary condition. We chose the case of $Re=3000$, $\alpha=1$ and $\beta=0$, which is a two-dimensional mode with zero spanwise velocity component. For the numerical simulation of the forward and backward linear systems, we used 64 grid points in the wall normal direction and a time-step size of $\Delta t$= 0.01. Typically 5$\sim$8 iterations are sufficient for achieving a converged value (with a threshold of $10^{-3}$) for the largest eigenvalue using the Krylov subspace method. Figure \ref{fig:validation}(a) shows the comparison of our result with the reference values and obviously the two sets agree well. Therefore, our method can accurately calculate the transient growth of small perturbations. Note that for unstable modes, $G(t)$ will exponentially grow at sufficiently large times. \BS{We also calculated the critical Reynolds numbers of $\beta=0$ modes for a few streamwise slip lengths using the velocity-vorticity formulation and compared with the reference values from Ghosh {\it et al.} 2014 \cite{Ghosh2014b}. Note that the Reynolds number definition is different in \cite{Ghosh2014b} and the values were converted to our definition for comparison. FIG. \ref{fig:validation}(b) shows that our method can accurately obtain the unstable eigenvalues.}

\subsection{Streamwise slip}

\begin{figure}
\centering
\includegraphics[width=0.4\linewidth]{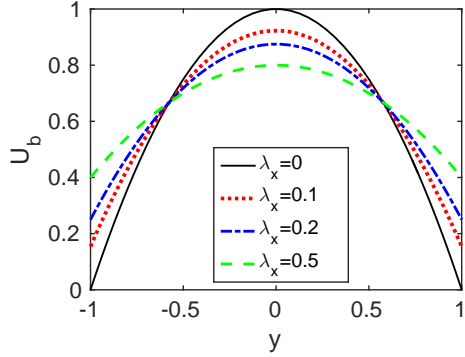}
\caption[base flow]{
    \label{fig:baseflow} 
     The velocity profiles of the base flow with $\lambda_x=0$, 0.1, 0.2 and 0.5} 
\end{figure}

Streamwise slip at the wall will change the base flow. Figure \ref{fig:baseflow} shows the analytical solutions of the the base flow with $\lambda_x=0$, 0.1, 0.2 and 0.5. The velocity profiles are still parabolas but with different boundary velocities. As $\lambda_x$ increases, the slip velocity at the wall increases. At the limit of $\lambda_x\to\infty$, the full slip boundary condition $\partial{u_x}/{\partial n}=0$, as for inviscid flow, is approached. For all cases, the total volume flux (or bulk speed) in the channel is fixed to the value for the no-slip case and the Reynolds number is therefore the same for all cases.

\subsubsection{Linear stability}
\label{sec:linear_instability_streamwise}
\begin{figure}
\includegraphics[width=0.8\linewidth]{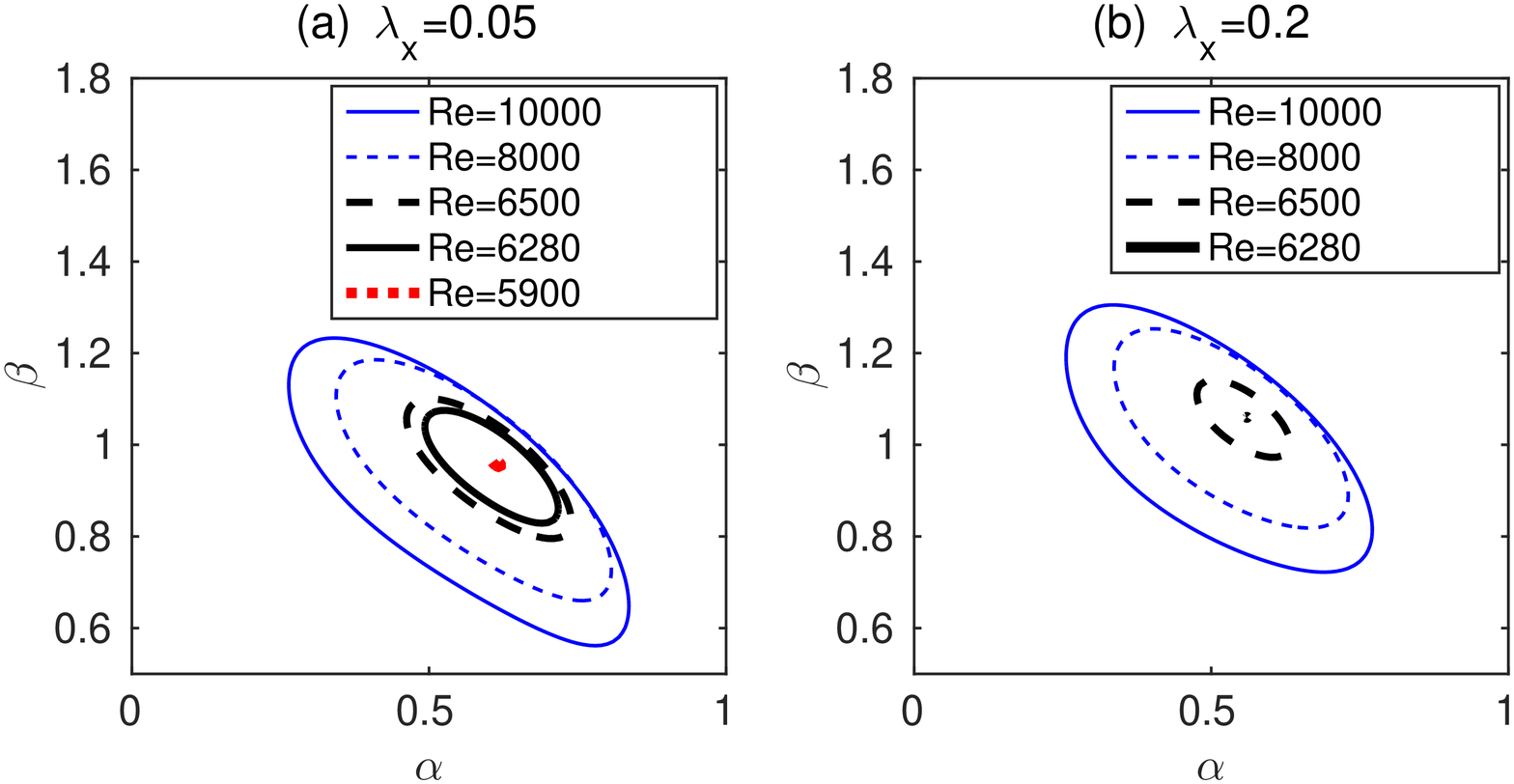}\\
\includegraphics[width=0.9\linewidth]{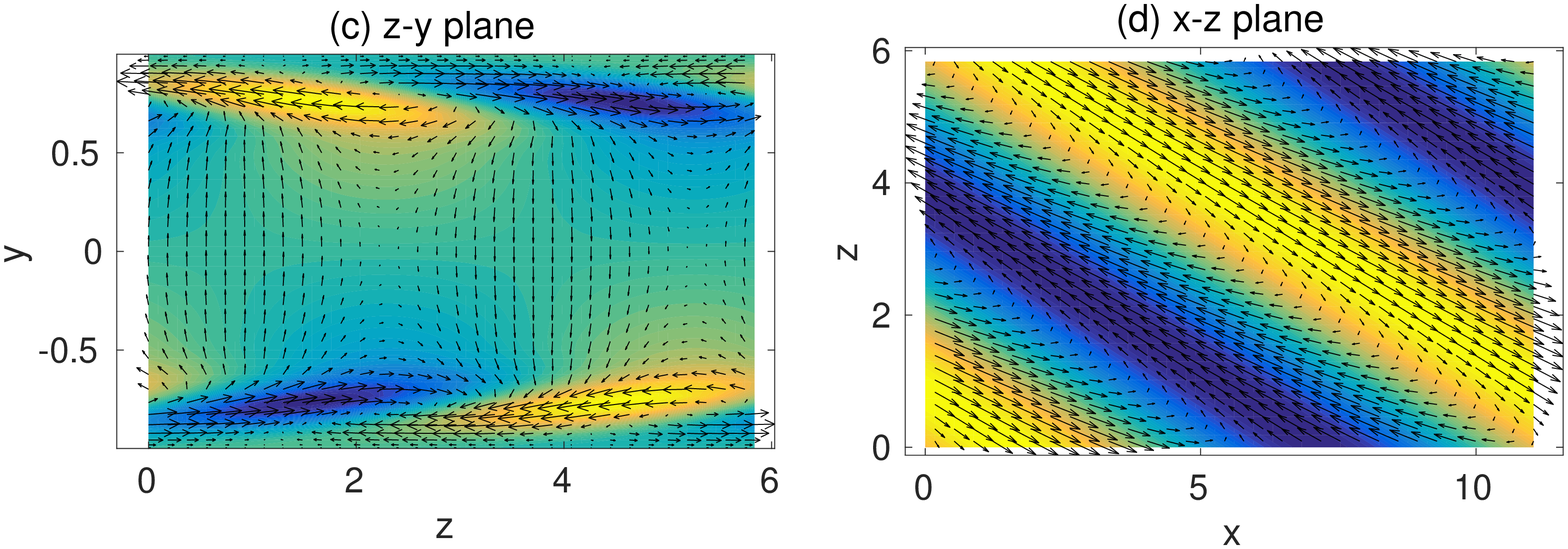}
\caption[unstable_region_streamwise_slip]{
    \label{fig:unstable_region_streamwise}
     The stability boundary for $\lambda_x=0.05$ (a) and 0.2 (b) in the $\alpha$-$\beta$ plane. The regions enclosed by the curves are linearly unstable regions. The leading unstable mode $(\alpha\simeq 0.56, \beta\simeq 1.06)$ of the $Re=6280$ and $\lambda_x=0.2$ case visualized in the $z$-$y$ plane at $x=0$ (c) and $x$-$z$ plane at y=0.8 (d). In panel (c) $u_x$ is plotted as the colormap with yellow representing positive and blue representing negative values with respect to the base flow. In (d) $u_y$ is plotted as the colormap.} 
\end{figure}

\begin{figure}
\includegraphics[width=0.5\linewidth]{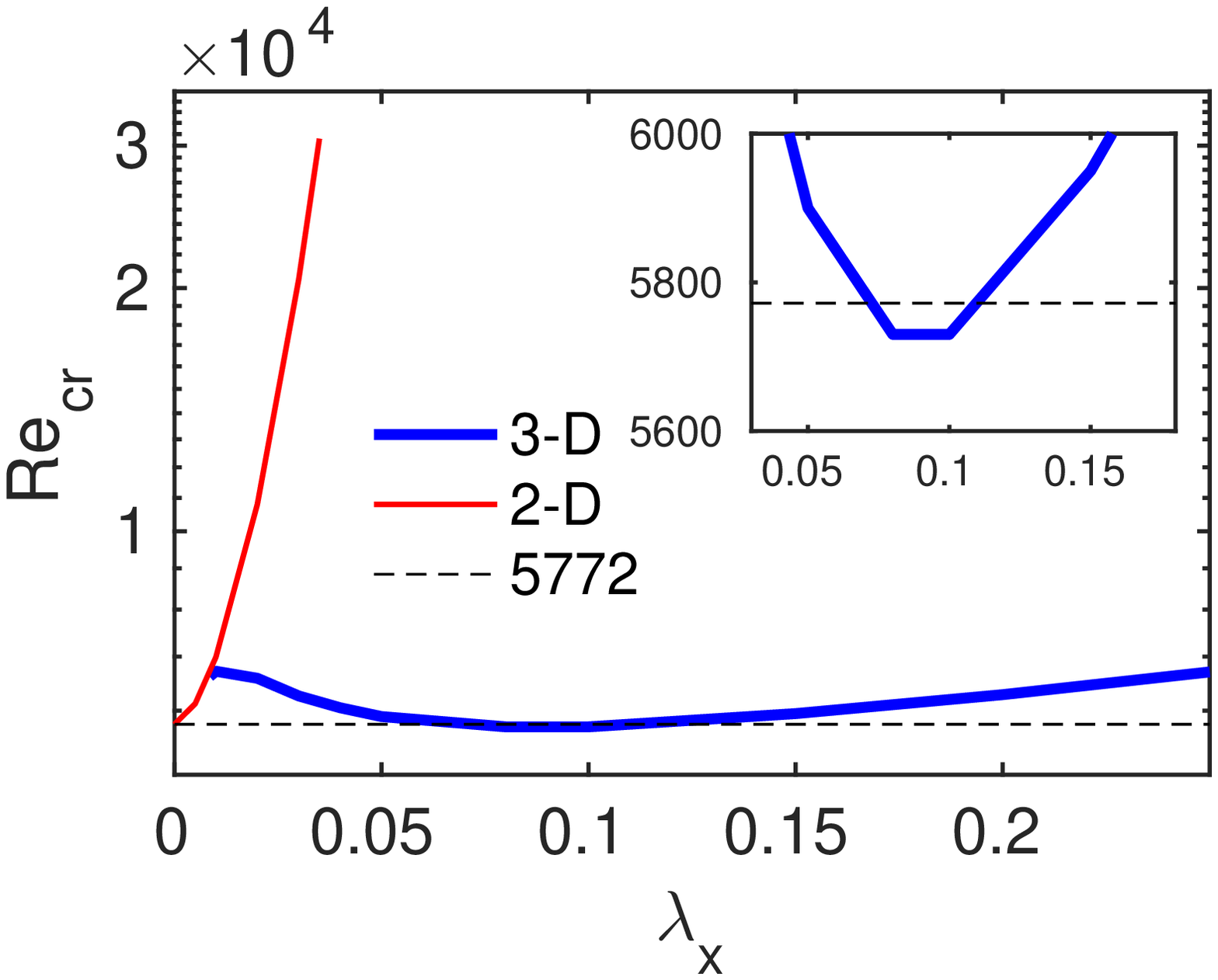}
\caption[critical_Re_streamwise_slip]{
    \label{fig:critical_Re_streamwise}
     \BS{The critical Reynolds number $Re_{cr}$ as a function of $\lambda_x$ for both 2-D modes (red thin line, see also FIG. \ref{fig:validation}(b)) and 3-D modes (blue bold line). $Re_{cr}=5772$ for no-slip channel flow is plotted as a dashed line for comparison. The inset is a close-up of the region $0.03<\lambda_x<0.18$.}} 
\end{figure}

Plane Poiseuille flow becomes linearly unstable above $Re_{cr}\simeq 5772$ and the leading unstable modes are spanwise invariant ($\beta=0$) modes, which are two-dimensional flows with zero spanwise velocity component. Interestingly, as shown in FIG. \ref{fig:unstable_region_streamwise}(a,b), streamwise slip can trigger three-dimensional (3-D) leading instabilities rather than two-dimensional ones. \BS{For a given slip length, the unstable region in the wave number plane shrinks as $Re$ decreases. The critical Reynolds number $Re_{cr}$, at which instability first occurs, can be searched by decreasing $Re$ step by step.} For examples, for $\lambda_x=0.05$, $Re_{cr}\simeq 5900$ and the leading unstable mode is the ($\alpha=0.61, \beta=0.96$) mode, and for $\lambda_x=0.2$, $Re_{cr}$ is a bit higher at 6280 and the leading unstable mode is ($\alpha=0.56, \beta=1.06$). Figure \ref{fig:unstable_region_streamwise}(c, d) visualize the detailed flow field of the latter case. In the $z$-$y$ plane, alternating high speed and low speed streaks arranged in the spanwise direction can be observed with rather complicated in-plane velocity field (see the vectors). In the $x$-$z$ plane at $y=0.8$, which cuts through the streaks, the flow exhibits straight structures tilted with respect to the streamwise direction.

\BS{To determine when 3-D leading instabilities set in as $\lambda_x$ increases, the critical Reynolds number $Re_{cr}$ as a function of $\lambda_x$ is calculated and shown in FIG. \ref{fig:critical_Re_streamwise} (the blue bold line). For comparison, $Re_{cr}$ of 2-D modes is also calculated (the red thin line). As $\lambda_x$ increases, $Re_{cr}$ associated with 2-D modes increases rapidly, as reported in previous studies \cite{Min2005, Lauga2005, Ghosh2014b}. We find that the leading instability is still 2-D below $\lambda_x\simeq 0.008$ but becomes 3-D at larger $\lambda_x$. This transition implies that, as slip length increases, the least stable modes have already switched from 2-D to 3-D ones before the system becomes linearly unstable. Above $\lambda_x\simeq 0.008$, $Re_{cr}$ does not undergo a monotonic increase, rather first decreases as $\lambda_x$ increases. Interestingly, it even drops below 5772 in the range $0.07\lesssim \lambda_x \lesssim 0.11$ (see the inset), indicating that streamwise slip even slightly destabilizes the flow in this small slip range. As $\lambda_x$ increases further, $Re_{cr}$ starts to increase but only exhibits a much slower growth compared to that of 2-D modes. This is consistent with that the unstable regions for $Re$=8000 and 10000 only shrink slightly as $\lambda_x$ increases from 0.05 to 0.2, see FIG. \ref{fig:unstable_region_streamwise}(a,b).} 

\subsubsection{Non-modal stability}
\begin{figure}
\centering
\includegraphics[width=0.9\linewidth]{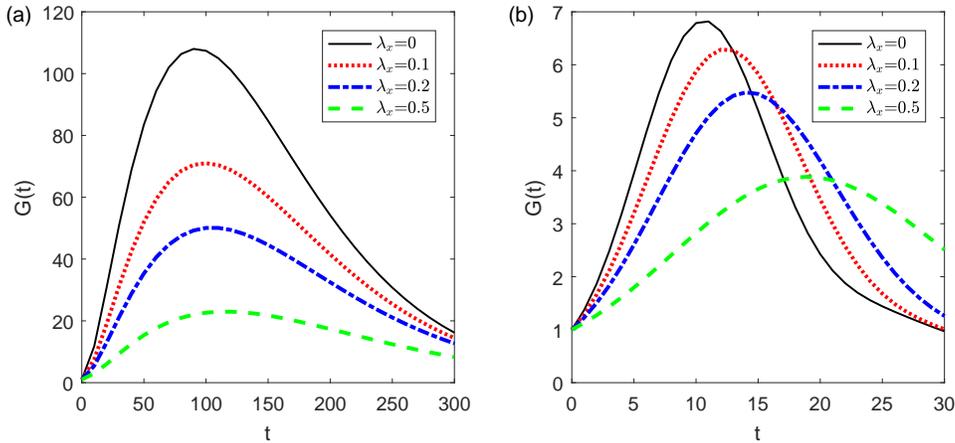}
\caption[transient growth Re=1000 with streamwise slip]{
    \label{fig:TG_streamwise_slip} 
     $G(t)$ of the 3-D mode $(\alpha=0, \beta=1)$ (a) and the 2-D mode $(\alpha=1, \beta=0)$ (b) at $Re=1000$ for $\lambda_x=0$, 0.1, 0.2 and 0.5. See the velocity profiles of the base flow in FIG. \ref{fig:baseflow}.} 
\end{figure}

In this section we investigate the non-modal stability of slip channel flow in the linearly stable regime. 
Figure \ref{fig:TG_streamwise_slip} shows the effects of streamwise slip on the non-modal transient growth of small perturbations. Clearly, streamwise slip reduces the transient growth and postpones the instant when the maximum transient growth $G_{\text max}:=\max_t G(t)$ is reached, hereafter referred to as $t_{\text max}$. This is expected because streamwise slip flattens the basic velocity profile (see FIG. \ref{fig:baseflow}) and therefore reduces the shear of the base flow all through the flow domain. The reduced shear results in weaker velocity streaks that streamwise vortices can generate by convecting the streamwise momentum in the radial direction. Therefore, the lift-up mechanism should be subdued. 
For instance, $G_{\text max}$ is reduced by more than a half with $\lambda_x=0.2$ and by about 80\% with $\lambda_x=0.5$ for the 3-D mode ($\alpha=0, \beta=1$). For the 2-D mode ($\alpha=1, \beta=0$), $G_{\text max}$ is relatively less reduced. For example, $G_{\text max}$ is reduced only by about 40\% even when $\lambda_x$ is increased to 0.5. Compared to the 3-D mode ($\alpha=1, \beta=0$), $t_{\text max}$ of this 2-D mode seems to be more sensitive to the slip: $t_{\text max}$ nearly doubles with $\lambda_x=0.5$, see FIG. \ref{fig:TG_streamwise_slip} (b). 
This is because, the transient growths of 2-D modes and 3-D modes come from different mechanisms. 
For 2-D modes, the transient growth mainly results from the Orr-mechanism, i.e., the amplification of disturbances initially tilted against the background shear until they are aligned with the shear under the distortion of the shear \cite{Jimenez2013}. Consequently, the growth occurs on a convection time scale given by the background shear. Therefore, the reduced background shear under streamwise slip will significantly enlarge the duration of the growth of 2-D modes, i.e., $t_{\text max}$. But for 3-D modes, the transient growth mainly results from the lift-up mechanism, i.e., the energy growth due to that long-lived streamwise vortices (rolls) continuously convect the streamwise momentum and generate strong streaks before they decay due to viscosity. As a consequence, the reduced background shear mainly reduces the magnitude of the streaks generated via the lift-up mechanism, i.e., $G_{\text max}$. 

\begin{figure}
\centering
\includegraphics[width=0.8\linewidth]{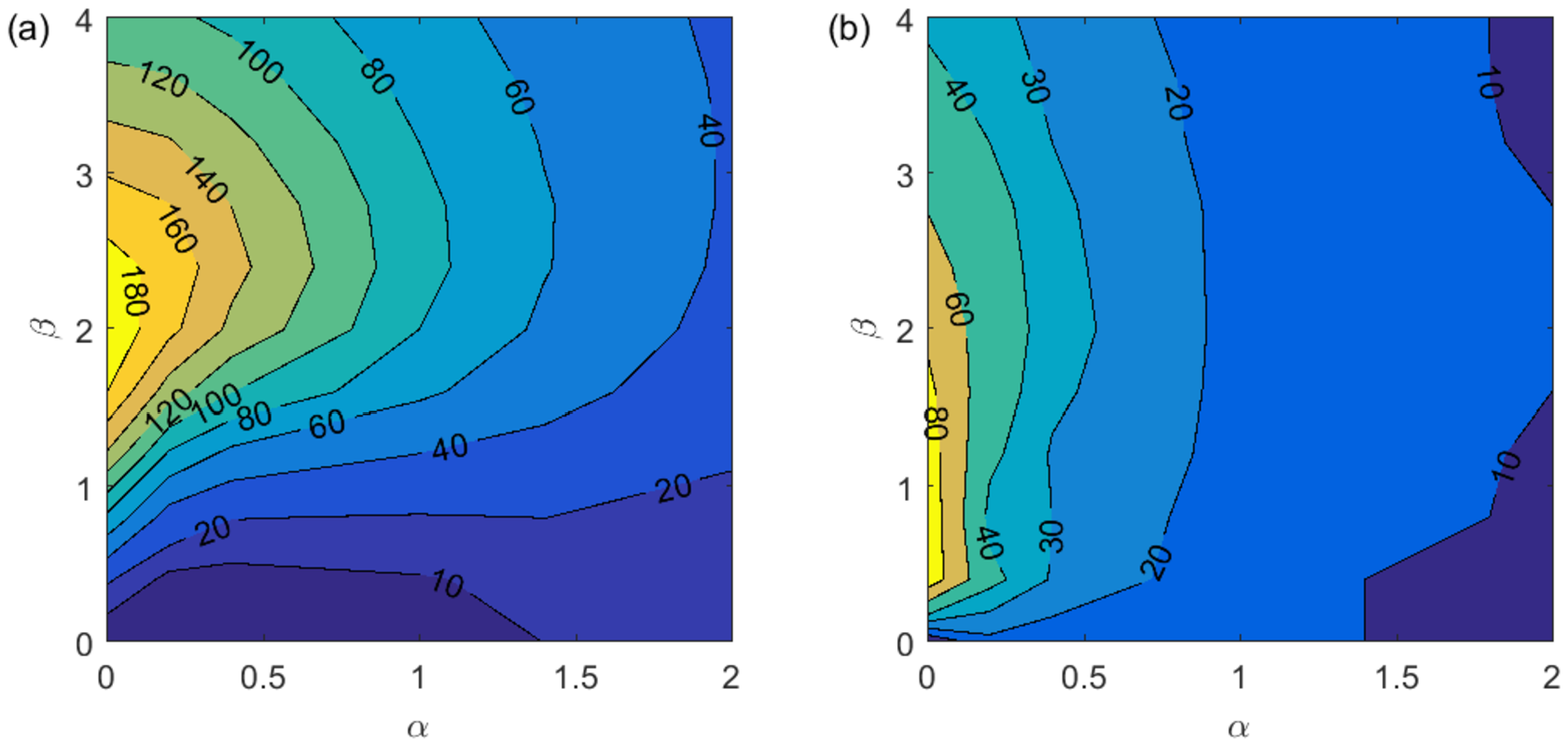}\\
\includegraphics[width=0.8\linewidth]{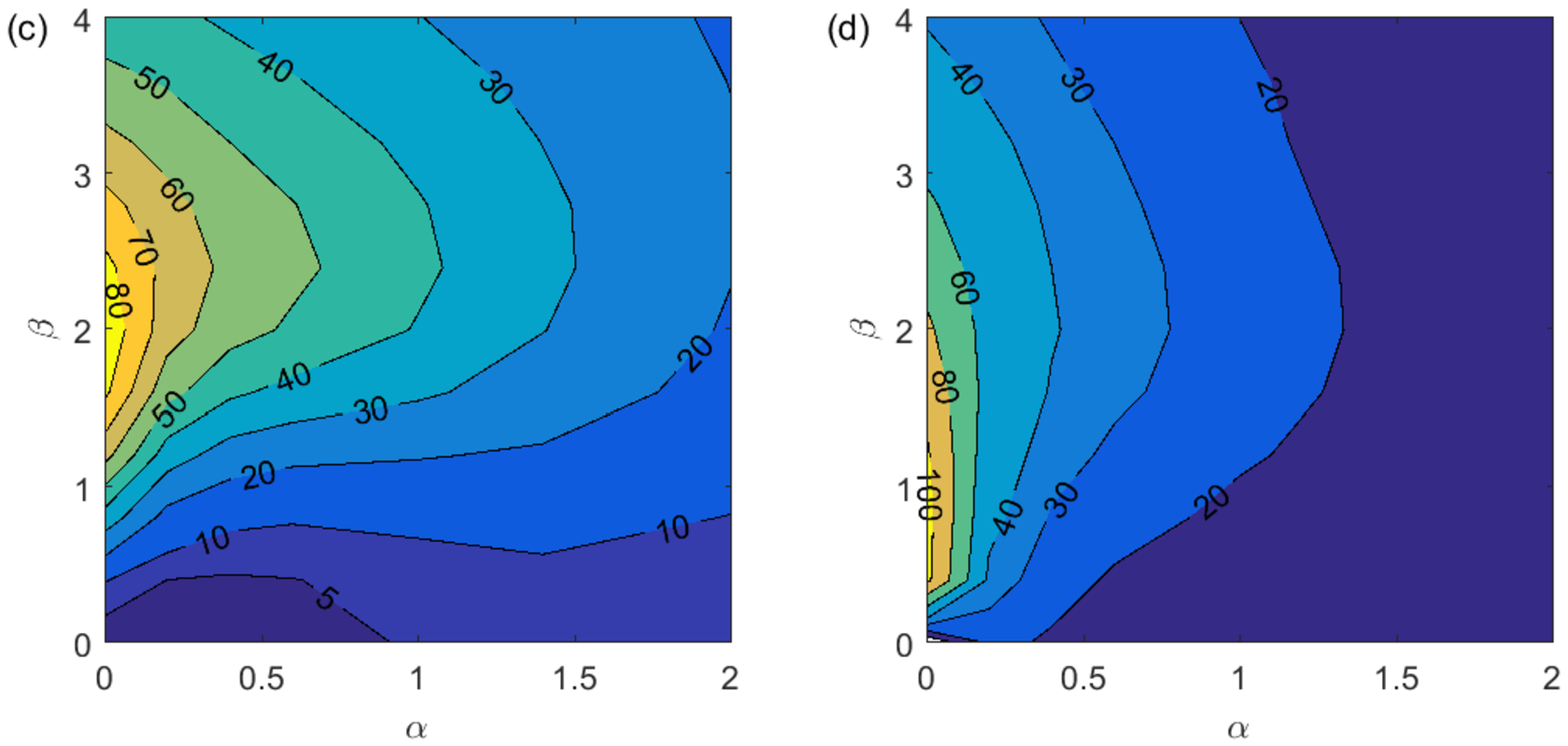}
\caption[transient growth and time contour Re=1000 lambdax=0.2]{
    \label{fig:TG_contour_streamwise_slip} 
     Contours in the wave number $\alpha-\beta$ plane of the maximum transient growth $G_{\text{max}}$ (a, c) and the corresponding $t_{\text{max}}$ (b, d) for $Re=1000$ with no-slip boundanry condition (a, b) and with streamwise slip length $\lambda_x=0.2$ (c, d).} 
\end{figure}

To show the effects of the slip on different modes, the contours of $G_{\text{max}}$ and $t_{\text{max}}$ for the $Re=1000$ and $\lambda_x=0.2$ case are plotted in the $\alpha$-$\beta$ wave number plane, see Fig. \ref{fig:TG_contour_streamwise_slip}. The results show that $G_{\text{max}}$ is reduced by roughly a factor of two compared to the no-slip case, for nearly all wave numbers considered in our study, whereas $t_{\text{max}}$ is not significantly affected, except for that of 2-D modes which is considerably enlarged by the reduced shear as discussed before.

However, in comparison with the no-slip case, no significant change can be observed in the distribution of both $G_{\text{max}}$ and $t_{\text{max}}$: $G_{\text{max}}$ still peaks at $\alpha=0$ and $\beta\simeq2$ and $t_{\text{max}}$ still peaks at $\alpha=0$ and $\beta\simeq 1$. This suggests that streamwise slip does not change the dominant flow structure during the transient growth stage of small perturbations.

\subsection{Spanwise slip}

Unlike streamwise slip, spanwise velocity slip does not affect the velocity profile of the base flow, i.e., the parabolic velocity profile $\bm U_b=(1-y^2)\bm e_x$ stays unchanged regardless of the value of the slip length and so does the background shear. 
\subsubsection{Linear stability}
\label{sec:linear_instability_spanwise}
\begin{figure}
\centering
\includegraphics[width=0.8\linewidth]{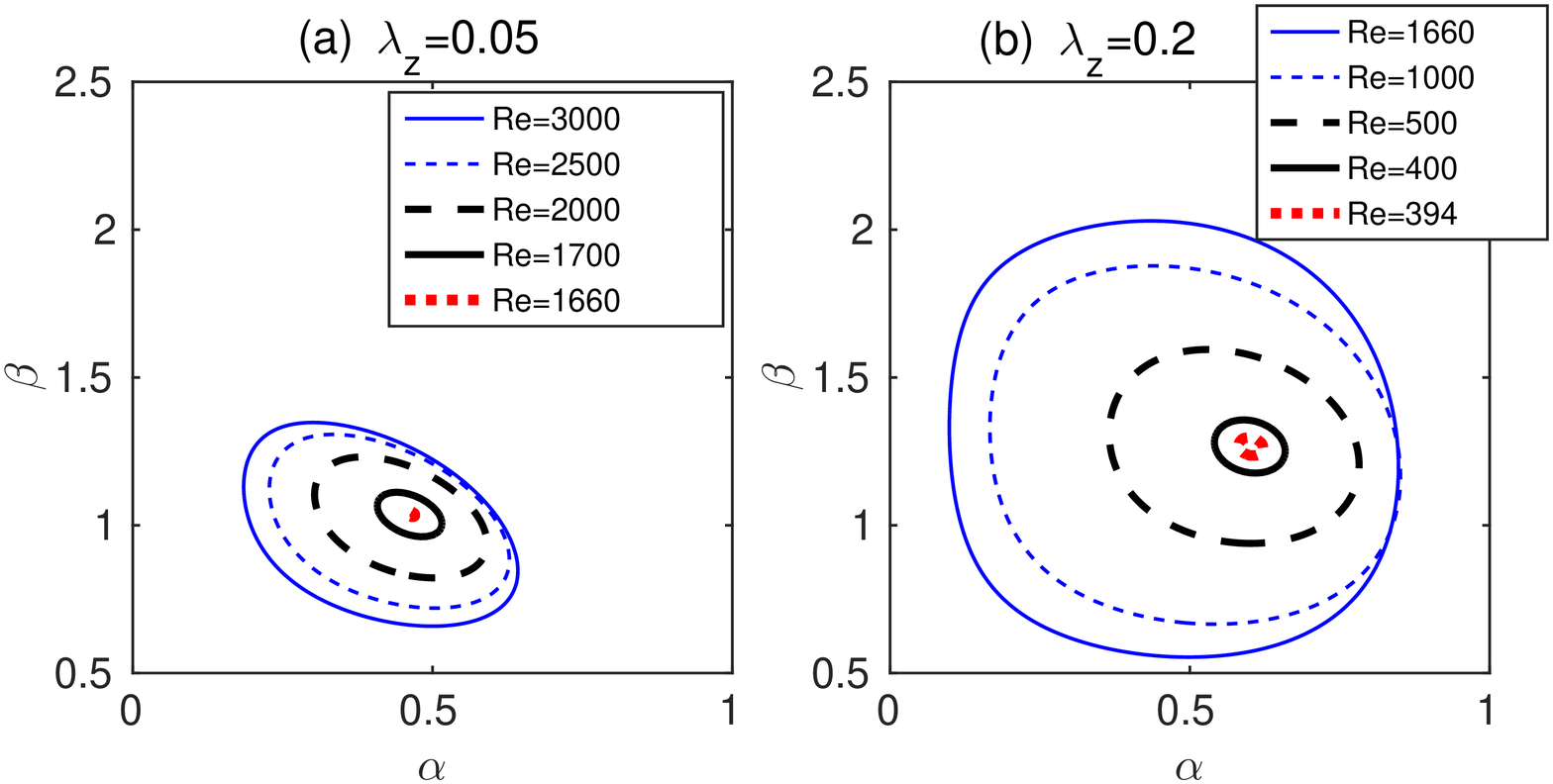}\\
\includegraphics[width=0.88\linewidth]{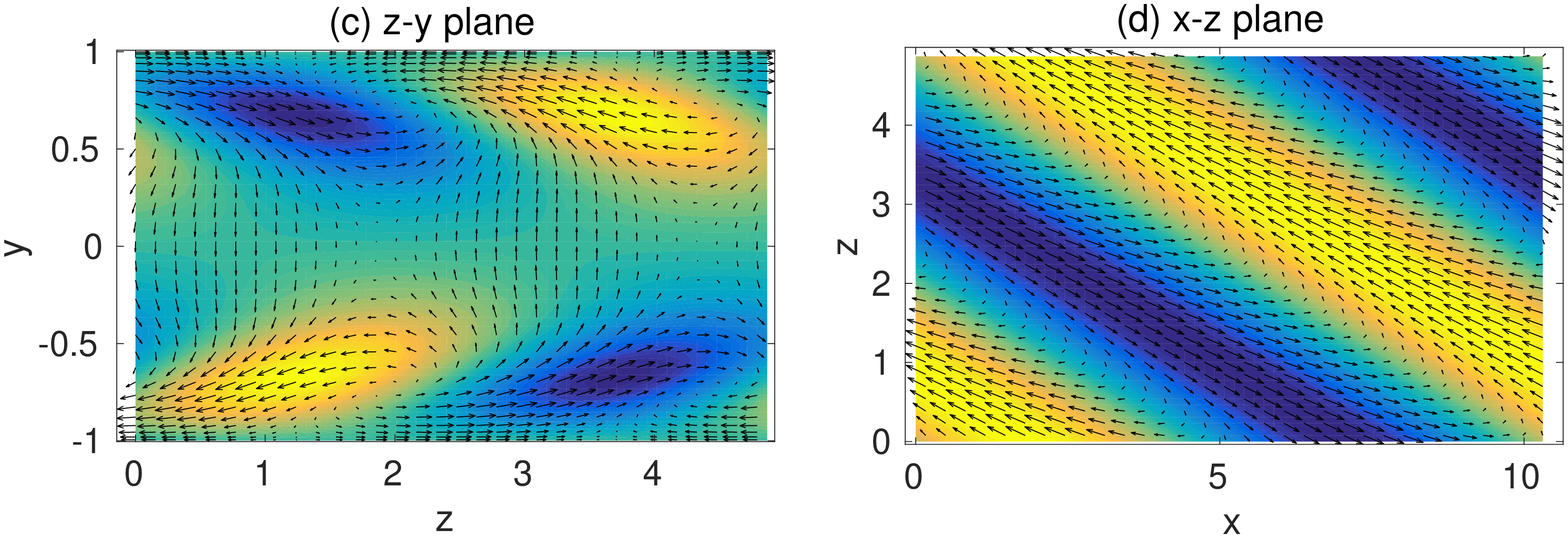}
\caption[unstable_region_spanwise_slip]{
    \label{fig:unstable_region_spanwise}
     The stability boundary for $\lambda_z=0.05$ (a) and $\lambda_z=0.2$ (b) in the $\alpha$-$\beta$ plane. The regions enclosed by the curves are linearly unstable regions. The leading unstable mode $(\alpha\simeq 0.6, \beta\simeq 1.27)$ of the $Re=394$ and $\lambda_z=0.2$ case visualized in the $z$-$y$ plane at $x$=0 (c) and $x$-$z$ plane at $y$=0.75 (d). In (c) $u_x$ is plotted as the colormap with yellow representing positive and blue representing negative values with respect to the base flow. In (d) $u_y$ is plotted as the colormap.} 
\end{figure}

Surprisingly, we find that spanwise slip can cause linear instability way below the critical Reynolds number $Re_{cr}$=5772 for the no-slip case, although the base flow is not affected by the slip. \BS{Similar to the streamwise slip case, spanwise slip also can trigger 3-D leading instabilities, see FIG.~\ref{fig:unstable_region_spanwise}(a, b). The figure also shows that the unstable region in the $\alpha$-$\beta$ wave number plane shrinks as $Re$ decreases for a given slip length. Similarly, we search for the critical Reynolds number by varying $Re$.} In our calculation, instability first occurs at $Re_{cr}\simeq 1660$ for $\lambda_z=0.05$ and at $Re_{cr}\simeq 394$ for $\lambda_z=0.2$. The leading unstable mode is $(\alpha\simeq 0.46, \beta\simeq 1.04)$ and $(\alpha\simeq 0.6, \beta\simeq 1.27)$, respectively. \BS{Besides, FIG.~\ref{fig:unstable_region_spanwise}(a, b) also show that the unstable region in the wave number plane expands rapidly as $\lambda_z$ increases, see the $Re=1660$ case.} The visualization of the flow field of the latter case is shown in FIG.~\ref{fig:unstable_region_spanwise}(c, d). It seems that the flow structure is very similar to that of the unstable mode in the streamwise slip case as shown in FIG. \ref{fig:unstable_region_streamwise}(c, d), exhibiting alternating high speed and low speed streaks that are tilted with respect to the streamwise direction.

However, for 2-D modes, we find that spanwise slip does not affect the instability, which still first occurs at $Re\simeq 5772$ with $\alpha\simeq1.02$ regardless of the value of $\lambda_z$. This is reasonable because 2-D modes are of zero spanwise velocity and therefore are not affected by the slip. 

\begin{figure}
\centering
\includegraphics[width=0.9\linewidth]{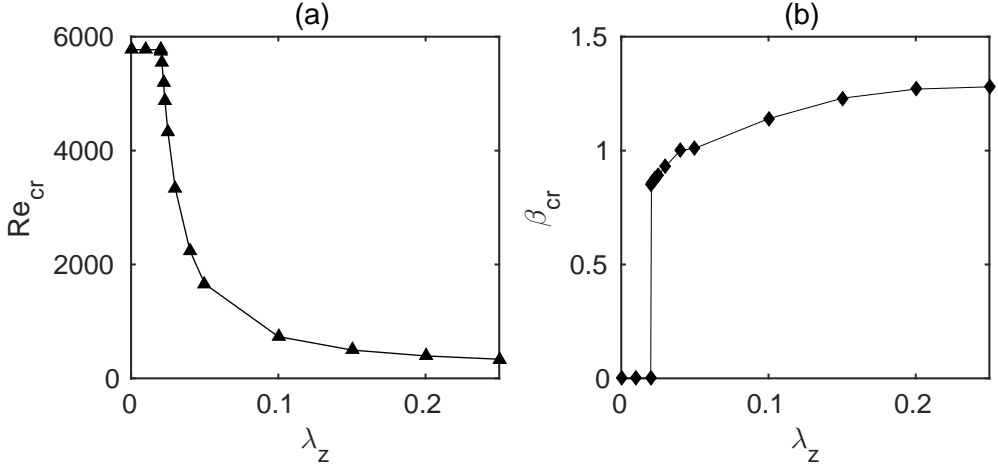}
\caption[critical_Re_spanwise_slip]{
    \label{fig:critical_Re_spanwise}
     \BS{The critical Reynolds number $Re_{cr}$ as a function of $\lambda_z$ (a) and the spanwise wavenumber associated with the leading unstable mode at the critical Reynolds number, $\beta_{cr}$, as a function of $\lambda_z$ (b).} }
\end{figure}

\BS{In order to find out the critical slip length for the appearance of 3-D leading instabilities, the critical Reynolds number $Re_{cr}$ is calculated as a function of $\lambda_z$ up to $\lambda_z=0.25$ and shown in FIG. \ref{fig:critical_Re_spanwise}(a). Our results show that the leading unstable modes stay as 2-D and are not affected by the spanwise slip up to $\lambda_z=0.02$, evidenced by the constant $Re_{cr}=5772$ and the constant corresponding spanwise wavenumber $\beta_{cr}=0$ (see panel (b)). As the slip increases further, the leading instability suddenly becomes 3-D and $Re_{cr}$ sharply drops. However, at large $\lambda_z$, $Re_{cr}$ only undergoes a slow decrease and reaches about 336 at $\lambda_z=0.25$. $\beta_{cr}$ as a function of $\lambda_z$ shows a jump at $\lambda_z=0.02$, indicating that, before instability sets in, 3-D modes have already become the least stable modes as $\lambda_z$ increases. This is the reason for the seemingly sudden appearance of 3-D leading instabilities far away from the $\beta$ axis in the $\alpha$-$\beta$ plane.}

\subsubsection{Non-modal stability}
\label{sec:nonmodal_spanwise}

\begin{figure}
\centering
\includegraphics[width=0.9\linewidth]{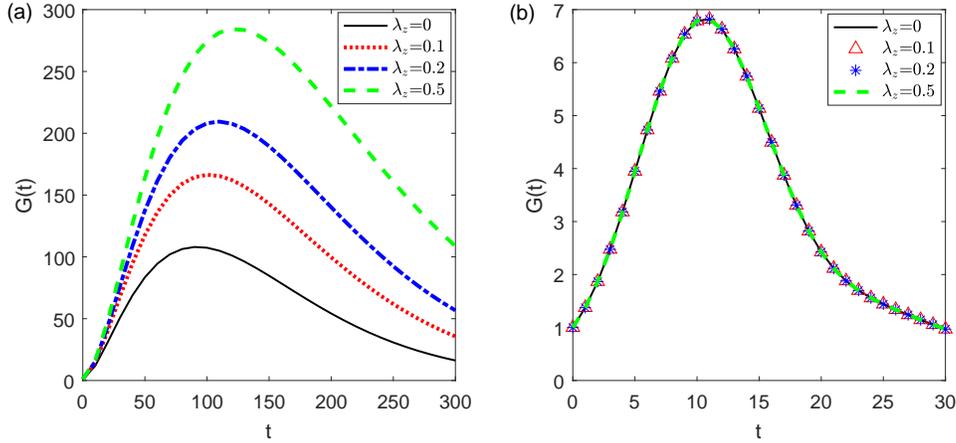}
\caption[transient growth Re=1000 with spanwise slip]{
    \label{fig:TG_spanwise_slip} 
     $G(t)$ of the 3-D mode $(\alpha=0, \beta=1)$ (a) and the 2-D mode $(\alpha=1, \beta=0)$ (b) at $Re=1000$ for spanwise slip lengths $\lambda_z=0$, 0.1, 0.2 and 0.5.} 
\end{figure}

The non-modal stability is also investigated in the linearly stable regime. Figure \ref{fig:TG_spanwise_slip} shows the transient growth as a function of time of the 3-D mode $(\alpha=0, \beta=1)$ at Re=1000 given different spanwise slip lengths. We can see that the slip results in larger transient growth. For instance, $G_{\text{max}}$ doubles with $\lambda_z=0.2$ and nearly triples with $\lambda_z=0.5$. On the other hand, it results in only slightly longer growth time window, i.e., $t_{\text{max}}$. In addition, it takes longer for perturbations to eventually decay due to viscosity when the slip is larger. In contrast, the transient growth of the 2-D mode $(\alpha=1, \beta=0)$ is not affected by the slip, regardless of the value of the slip length. Figure \ref{fig:TG_spanwise_slip} (b) clearly shows that G(t) for different $\lambda_z$'s exactly coincide. This is reasonable because for 2-D modes, the working amplification mechanism is the Orr-mechanism, which amplifies the tilted disturbances by shearing them and the growth solely depends on the background shear. In fact, for 2-D modes, the spanwise velocity component is zero and the slip boundary condition actually does not cause velocity slip in the spanwise direction.  

\begin{figure}
\centering
\includegraphics[width=0.8\linewidth]{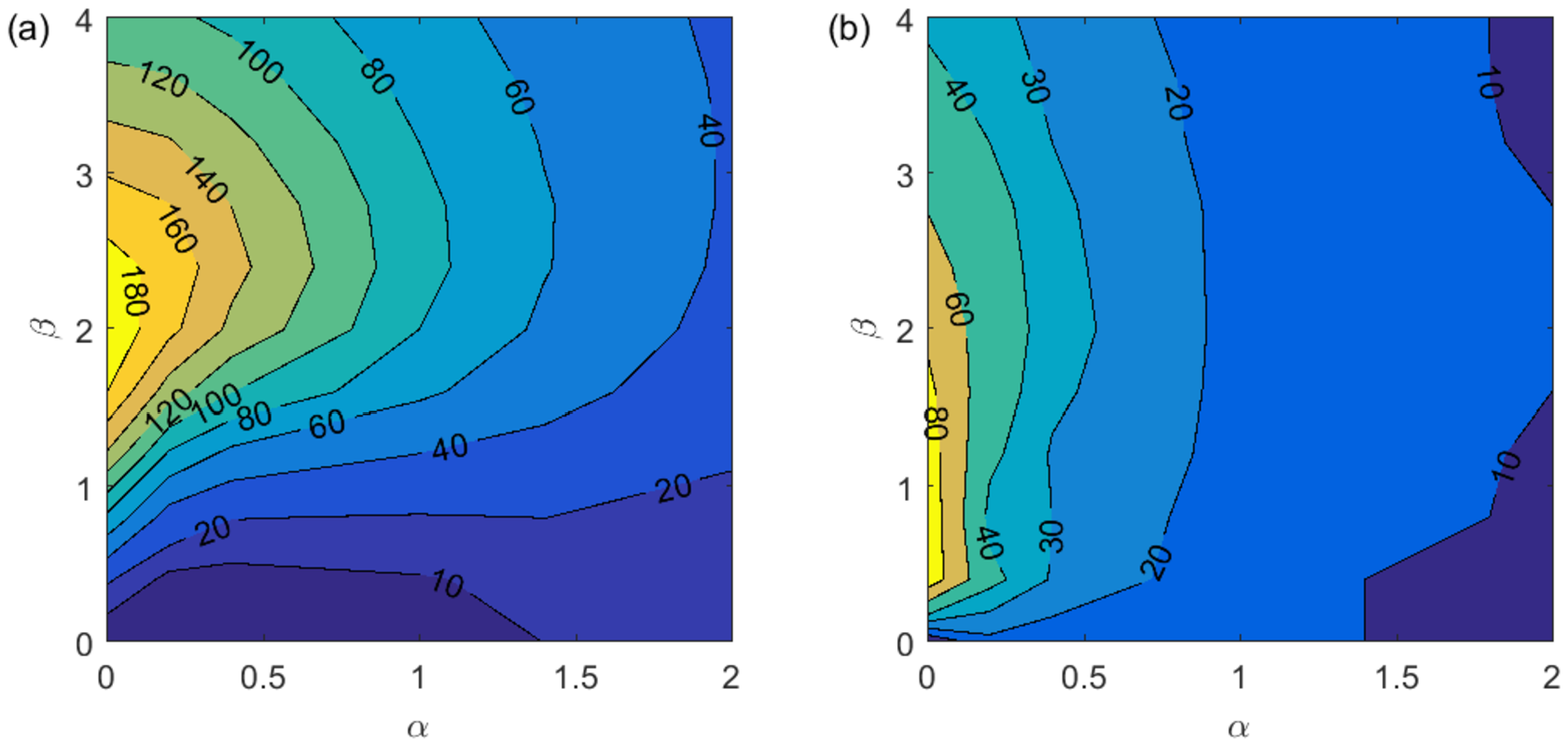}\\
\includegraphics[width=0.8\linewidth]{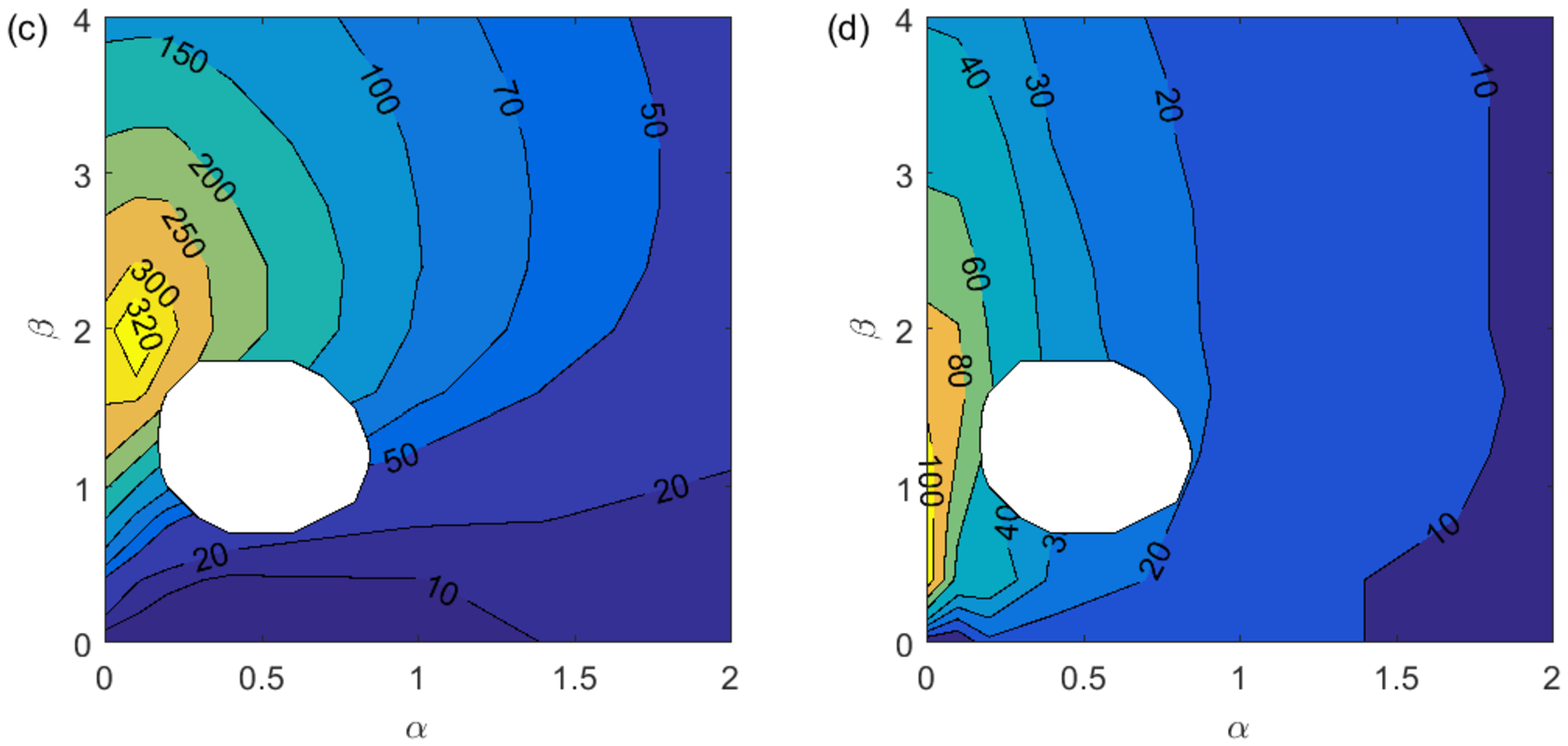}
\caption[transient growth and time contour Re=1000 with spanwise slip]{
    \label{fig:TG_contour_spanwise_slip}
    Contours in the wave number $\alpha-\beta$ plane of $G_{\text{max}}$ (a, c) and $t_{\text{max}}$ (b, d) for $Re=1000$. (a, b) No-slip case. (c, d) $\lambda_z=0.2$. The blank areas represent linearly unstable regions.} 
\end{figure}

\begin{figure}
\centering
\includegraphics[width=0.4\linewidth]{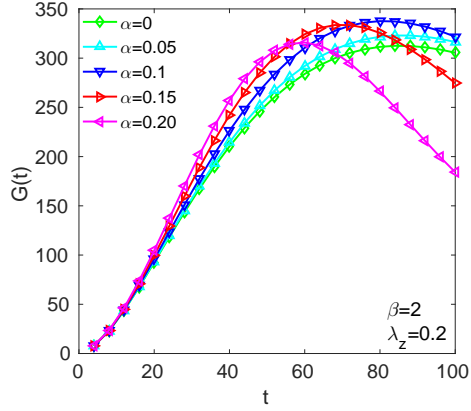}
\caption[transient growth and time contour Re=1000 with spanwise slip]{
    \label{fig:TG_Re1000_beta2_multiple_alpha}
   The influence of spanwise slip on the most amplified perturbations for $Re=1000$ and $\lambda_z=0.2$. Transient growth $G(t)$ for $\beta=2$ with $\alpha=0$, 0.05, 0.1, 0.15 and 0.2 are plotted.} 
\end{figure}

\begin{figure}
\centering
\includegraphics[width=0.7\linewidth]{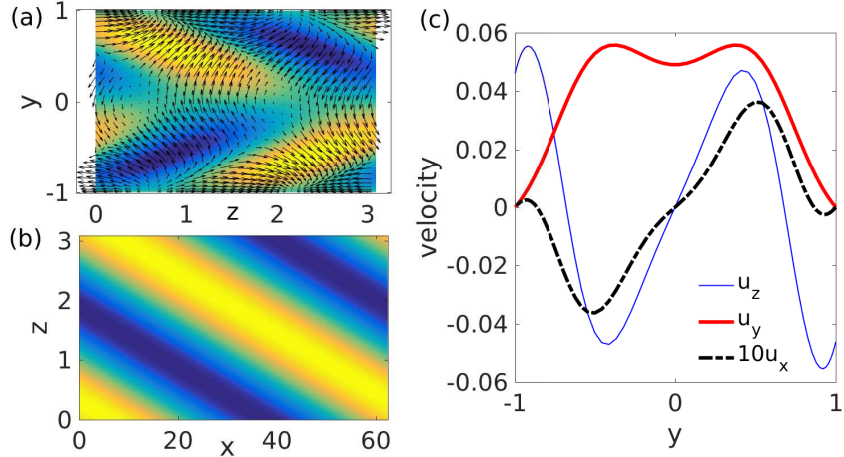}
\caption[transient growth and time contour Re=1000 with spanwise slip]{
    \label{fig:TG_Re1000_beta2_alpha0.1}
    The optimal perturbation of the mode ($\alpha=0.1$, $\beta=2$) for $Re=1000$ and $\lambda_z=0.2$, corresponding to the blue-down triangle curve in FIG. \ref{fig:TG_Re1000_beta2_multiple_alpha}. (a) Velocity field on the $z$-$y$ cross-section at $x=40$. Arrows represent the in-plane velocities and the colormap represents $u_x$. Blue color marks low speed regions and yellow high speed regions. (b) Contours of $u_x$ on the $x$-$z$ plane of the channel at $y=0.5$. (c) The velocity profiles of $u_y$, $u_z$ and $10u_x$ at the position $(x,z)=(0,0)$} 
\end{figure}

Figure \ref{fig:TG_contour_spanwise_slip} shows the contours of $G_{\text{max}}$ and $t_{\text{max}}$ for $Re=1000$ in the $\alpha$-$\beta$ wave number plane. Comparison is made between the no-slip (a, b) and $\lambda_z=0.2$ (c,d) cases. For 3-D ($\beta\neq 0$) modes, spanwise slip enlarges the transient growth and triggers a linearly unstable region. But in most part of the linearly stable region the patterns of the distribution of $G_{\text max}$ and $t_{\text max}$ in the two cases are roughly the same. 
However, a notable difference is that $G_{\text max} $ peaks at a small $\alpha$ value (about 0.1) rather than $\alpha=0$ as in the no-slip and streamwise slip cases. This indicates that the maximally amplified perturbations are no longer streamwise invariant ($\alpha=0$) modes (see FIG. \ref{fig:TG_contour_streamwise_slip}) but those with small finite streamwise wave numbers (long wavelengths). Figure \ref{fig:TG_Re1000_beta2_multiple_alpha} shows the transient growth of the $\beta=2$ modes with a few small $\alpha$'s for $\lambda_z=0.2$. The most amplified mode has an axial wave number of about $\alpha=0.1$, with $G_{\text max}$ roughly 6\% higher than that of the $\alpha=0$ mode (see the comparison between the green-diamond and blue-down triangle curves). The optimal perturbation for ($\alpha=0.1$, $\beta=2$) is visualized in FIG. \ref{fig:TG_Re1000_beta2_alpha0.1}. The velocity field plotted on the $z$-$y$ cross-section of the channel (panel (a)) shows that the flow does not feature streamwise rolls as in the no-slip case. Nevertheless, lift-up of low speed flow towards the channel center by the wall-normal velocity component appears to be dominant, suggesting that the dominant growth mechanism is still the lift-up mechanism.  
Besides, panel (b) shows the structure of the flow on the $x$-$z$ cut-plane at $y=0.5$, which presents straight flow structures tilted with respect to the streamwise direction by a small angle, similar to the linearly unstable modes shown in FIG. \ref{fig:unstable_region_streamwise} and \ref{fig:unstable_region_spanwise} but with much larger streamwise wavelength. Panel (c) plots the velocity profiles of $u_z$, $u_y$ and $u_x$ at the position $(x,z)$=(0, 0), which clearly show a central symmetry for $u_z$ and $u_x$ and a mirror symmetry for $u_y$ about the channel mid-plane. Note that the streamwise velocity of this optimal perturbation is one order of magnitude smaller than the other two components and its value in the figure is scaled by a factor of 10.

Regarding $t_{\text max}$, except for the linearly unstable region caused by spanwise slip, there is no significant change in either the magnitude or distribution for both 2-D and 3-D modes. This indicates that spanwise slip does not significantly affect the transient growth time window of small perturbations. 

\subsection{Isotropic slip}
With equal slip in streamwise and spanwise directions (we refer to as isotropic slip), Lauga \& Cossu (2005) \cite{Lauga2005} have investigated the modal instability of 2-D modes for small values of slip length up to 0.03 and showed that the slip greatly suppresses the instability. They showed that the critical Reynolds number of 2-D modes increases to nearly 20000 with $\lambda_x=\lambda_z=0.03$. We extend the slip to larger values and find that instability first occurs at $Re\simeq 1.85\times 10^5$ for $\lambda_x=\lambda_z=0.05$ and the most unstable mode is still 2-D. This sharp increase of the critical Reynolds number indicates a strong sensitivity of the instability on the slip length, agreeing with previous studies \cite{Lauga2005}. Surprisingly, the 3-D instabilities in the pure streamwise and spanwise cases (see FIG. \ref{fig:unstable_region_streamwise} and \ref{fig:unstable_region_spanwise}) do not occur in the presence of isotropic slip.

\begin{figure}[!ht]
\centering
\includegraphics[width=0.9\linewidth]{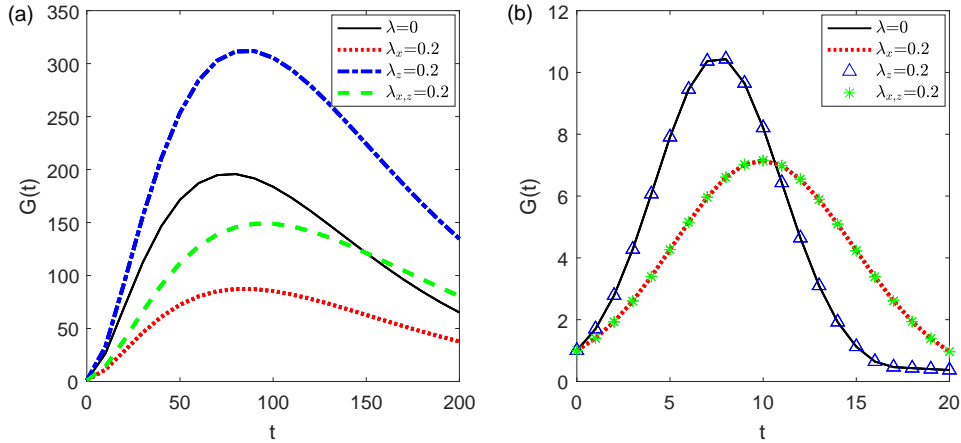}
\caption[Transient growth with isotropic slip]{
    \label{fig:mixed_slip}
     $G(t)$ of the 3-D mode ($\alpha=0$, $\beta=2$) (a) and the 2-D mode ($\alpha=2$, $\beta=0$) (b) for Re=1000 with isotropic slip of $\lambda_x=\lambda_z=0.2$ (green dashed in (a) and green star in (b)). For comparison, the no-slip (black solid) case, the pure streamwise slip case with $\lambda_x=0.2$ (red dotted) and the pure spanwise slip case with $\lambda_z=0.2$ (blue dash-dotted in (a) and triangle in (b)) are also plotted.} 
\end{figure}

Lauga \& Cossu (2005) \cite{Lauga2005} and Min \& Kim (2005) \cite{Min2005} showed that the non-modal transient growth is only modestly affected by the slip. Here we only briefly compare the non-modal transient growth of the isotropic slip case with the pure streamwise and spanwise slip cases and investigate the slip in which direction has the dominant effect. Figure \ref{fig:mixed_slip} shows the transient growth of the 3-D mode ($\alpha=0$, $\beta=2$) and the 2-D mode ($\alpha=2$, $\beta=0$) as a function of time for various slip settings. For the 3-D mode, the transient growth is smaller and slower (the green-dashed line) compared to the no-slip case (the black solid line). At large times, however, the transient growth is larger and decreases more slowly than the no-slip case. Comparing to the pure streamwise slip (the red dotted line) and pure spanwise slip (the blue dash-dotted line) cases, the trend seems to suggest that, at least for this (nearly) optimally amplified mode, $G_{\text max}$ is dominated by streamwise slip  whereas $t_{\text max}$ and the decay at large times are more strongly influenced by spanwise slip. Whereas, for the 2-D mode, the results for the isotropic slip and pure streamwise slip cases are identical, indicating that the transient growth is solely determined by the streamwise velocity slip. This is consistent with the analysis in section \ref{sec:nonmodal_spanwise} that spanwise slip does not affect the transient growth of 2-D modes.  

\section{Discussion and conclusion} 

\BS{In this work, we studied the stability of channel flow with large slip lengths considering the increasingly large effective slip length achieved in experiments \cite{Lee2008,Lee2009}. We found that streamwise slip suppresses the instability at small slip length and the most unstable modes are still 2-D modes, in agreement with previous findings \cite{Min2005,Lauga2005,Ghosh2014b}. However, as $\lambda_x$ is above about 0.008, 3-D instabilities cut in at lower $Re$ compared to 2-D ones. $Re_{cr}$ first slightly decreases and then modestly increases as $\lambda_x$ increases. It only reaches 6700 at $\lambda_x=0.25$, the largest slip length considered in our study. Overall, instability is only slightly suppressed in terms of critical Reynolds number. Interestingly, streamwise slip even destabilizes the flow in a small slip length range between 0.07 and 0.11, with $Re_{cr}$ slightly below 5772. 
Similarly, at small slip length, spanwise slip does not affect the instability. However, above $\lambda_z=0.02$, 3-D leading instabilities appear and $Re_{cr}$ sharply decreases as the slip increases. $Re_{cr}$ is lower than 5772 by more than an order of magnitude at large slip lengths and the trend shows a monotonic and slow decease if $\lambda_z$ increases further. We investigated the flow field of the 3-D unstable modes for both streamwise and spanwise slip and found that the flow manifests straight flow structures tilted with respect to the streamwise direction.}
These three-dimensional instabilities were not reported in former studies \cite{Lauga2005, Min2005,Ghosh2014b} seemingly because the instability was only studied for 2-D modes and as a result, the suppression of the instability by streamwise slip was overestimated. Nevertheless, 3-D leading instabilities and the dual effect of the slip were observed in multi-fluid channel flows with slip boundary condition \cite{Chattopadhyay2018,Ghosh2014a,Ghosh2014b,Ghosh2015,Sahu2011}.

Streamwise slip was shown to reduce the transient growth of all perturbations because it subdues the lift-up mechanism by reducing the background shear, in agreement with previous studies \cite{Lauga2005, Min2005}. However, it does not change the distribution of the maximal transient growth $G_{\text max}$ in the wave number plane, indicating that it will not change the dominant flow structures during the transient growth process of small perturbations. In our study, the optimal perturbations are still streamwise rolls with $\alpha=0$ and $\beta\simeq 2$, which agrees with the findings of \cite{Lauga2005, Min2005}. Streamwise slip slightly enlarges the growth time window $t_{\text max}$ of 3-D modes and significantly enlarges that of 2-D modes. In contrast, spanwise slip does not affect the base flow, therefore, the background shear remains unchanged regardless of the spanwise slip length. However, our results showed that it enlarges the transient growth of small disturbances compared to the no-slip case, in agreement with \cite{Min2005}. Interestingly, as the spanwise slip length increases, the distribution of the maximum transient growth in the wave number plane changes, with $G_{\text max}$ peaking at small finite $\alpha$ instead of $\alpha=0$ as in the no-slip, streamwise slip and small spanwise slip cases \cite{Min2005}. Therefore, in the presence of large spanwise slip, the most amplified flow structures during the transient growth process will be long-streamwise wavelength stuctures tilted with respect to the steamwise direction, instead of streamwise rolls. 

When equal slip length in streamwise and spanwise directions present, we found that the three-dimensional leading instabilities that would occur in pure streamwise and spanwise slip cases are removed, therefore, the linear instability is greatly suppressed. Besides, the non-modal transient growth is dominated by the streamwise slip and consequently is also suppressed. Hence, earlier instability and larger transient growth can only be triggered by introducing anisotropy in the velocity slip with large spanwise slip but small streamwise slip, which may be interesting for some applications that require increasing mixing rate of mixtures. Besides, it will be interesting to study the transition to turbulence when modal and non-modal growth mechanisms both exist and cause comparable energy growth, which is our ongoing work.

\subsection{Acknowledgements}
The authors acknowledge the financial support from Tianjin University under grant number 2018XZ-0006.
 
%{\footnotesize\begin{thebibliography}{88}
\bibliographystyle{unsrt}

\begin{thebibliography}{10}

\bibitem{Tao2018}
J.~J. Tao, B.~Eckhardt, and X.~M. Xiong.
\newblock Extended localized structures and the onset of turbulence in channel
  flow.
\newblock {\em Physical Review Fluids}, 3:011902, 2018.

\bibitem{Trefethen1993}
L.~N. Trefethen, A.~E. Trefethen, S.~C. Reddy, and T.~A. Driscoll.
\newblock Hydrodynamic instability without eigenvalues.
\newblock {\em Science}, 261:578--584, 1993.

\bibitem{Reddy1993}
S.~Reddy and D.~S. Henningson.
\newblock Energy growth in viscous channel flows.
\newblock {\em J.\,Fluid Mech.}, 252:209--238, 1993.

\bibitem{Schmid1994}
P.~J Schmid and D.~S. Henningson.
\newblock Optimal energy density growth in hagen-poiseuille flow.
\newblock {\em J.\,Fluid Mech.}, 277:197--225, 1994.

\bibitem{Schmid2007}
P.~J Schmid.
\newblock Nonmodal stability theory.
\newblock {\em Ann.\ Rev.\ Fluid Mech.}, 39:129--162, 2007.

\bibitem{Lee2008}
C.~Lee, C.-H. Choi, and C.-J Jim.
\newblock Structured surfaces for a giant liquid slip.
\newblock {\em Phys.\ Rev.\ Lett.}, 101:064501, 2008.

\bibitem{Lee2009}
C.~Lee and C.-J Jim.
\newblock Maximizing the giant liquid slip on superhydrophobic microstructures
  by nanostructuring their sidewalls.
\newblock {\em Langmuir}, 25:12812, 2009.

\bibitem{Voronov2008}
R.~S. Voronov, D.~V. Papavassiliou, and L.~L. Lee.
\newblock Review of fluid slip over superhy{-}drophobic surfaces and its
  dependence on the contact angle.
\newblock {\em Industr. Eng. Chem. Res.}, 47:2455--2477, 2008.

\bibitem{Chattopadhyay2018}
G.~Chattopadhyay, K.~C. Sahu, and R.~Usha.
\newblock Spatio{-}temporal instability of two superposed fluids in a channel
  with boundary slip.
\newblock {\em International Journal of Multiphase Flow}, 113:264--278, 2019.

\bibitem{Lauga2003}
E.~Lauga and H.~A. Stone.
\newblock Effective slip in pressure-driven stokes flow.
\newblock {\em J.\,Fluid Mech.}, 489:55--77, 2003.

\bibitem{Bazant2008}
M.~Z. Bazant and O.~I. Vinogradova.
\newblock Tensorial hydrodynamic slip.
\newblock {\em J.\,Fluid Mech.}, 613:125--134, 2008.

\bibitem{kamrin2011}
K.~Kamrin, M.~Z. Bazant, and H.~A. Stone.
\newblock Effective slip boundary conditions for arbitrarily patterned
  surfaces.
\newblock {\em Phys.\ Fluids}, 23:031701, 2011.

\bibitem{Park2013}
H.~Park, H.~Park, and J.~Kim.
\newblock A numerical study of the effects of superhydrophobic surface on
  skin{-}friction drag in turbulent channel flow.
\newblock {\em Phys.\ Fluids}, 25:110815, 2013.

\bibitem{Seo2016}
J.~Seo and A.~Mani.
\newblock On the scaling of the slip velocity in turbulent flows over
  superhydrophobic surfaces.
\newblock {\em Phys.\ Fluids}, 28:025110, 2016.

\bibitem{Yu2016}
K.~H. Yu, C.~J. Teo, and B.~C. Khoo.
\newblock Linear stability of pressure{-}driven flow over longitudinal
  superhydrophobic grooves.
\newblock {\em Phys.\ Fluids}, 28:022001, 2016.

\bibitem{Kumar2016}
A.~Kumar, S.~Datta, and D.~Kalyanasundaram.
\newblock Permeability and effective slip in confined flows transverse to wall
  slippage patterns.
\newblock {\em Phys.\ Fluids}, 28:082002, 2016.

\bibitem{Aghdam2016}
S.~H. Aghdam and P.~Ricco.
\newblock Laminar and turbulent flows over hydrophobic surfaces with
  shear{-}dependent slip length.
\newblock {\em Phys.\ Fluids}, 28:035109, 2016.

\bibitem{Gersting1974}
Jr. J.~M. Gersting.
\newblock Hydrodynamic instability of plane porous slip flow.
\newblock {\em Phys.\ Fluids}, 17:2126--2127, 1974.

\bibitem{Vinogradova1999}
O.~I. Vinogradova.
\newblock Slippage of water over hydrophobic surfaces.
\newblock {\em Int. J. Min. Process}, 56:31--60, 1999.

\bibitem{Chu2004}
A.~Kwang-Hua Chu.
\newblock Instability of navier slip flow of liquids.
\newblock {\em C. R. Mechanique}, 332:895--900, 2004.

\bibitem{Lauga2005}
E.~Lauga and C.~Cossu.
\newblock A note on the stability of slip channel flows.
\newblock {\em Phys.\ Fluids}, 17:088106, 2005.

\bibitem{Min2005}
T.~Min and J.~Kim.
\newblock Effects of hydrophobic surface on stability and transition.
\newblock {\em Phys.\ Fluids}, 17:108106, 2005.

\bibitem{Ghosh2014a}
S.~Ghosh, R.~Usha, and K.~C. Sahu.
\newblock Double{-}diffusive two{-}fluid flow in a slippery channel: A linear
  stability analysis.
\newblock {\em Phys.\ Fluids}, 26:127101, 2014.

\bibitem{Ghosh2014b}
S.~Ghosh, R.~Usha, and K.~C. Sahu.
\newblock Linear stability analysis of miscible two-fluid flow in a channel
  with velocity slip at the walls.
\newblock {\em Phys.\ Fluids}, 26:014107, 2014.

\bibitem{Ghosh2015}
S.~Ghosh, R.~Usha, and K.~C. Sahu.
\newblock Absolute and convective instabilities in double{-}diffusive two-fluid
  flow in a slippery channel.
\newblock {\em Chemical Engineering Science}, 134:1--11, 2015.

\bibitem{Chattopadhyay2017}
G.~Chattopadhyay, R.~Usha, and K.~C. Sahu.
\newblock Core{-}annular miscible two{-}fluid flow in a slippery pipe: A
  stability analysis.
\newblock {\em Phys.\ Fluids}, 29:097106, 2017.

\bibitem{Chakraborty2019}
S.~Chakraborty, T.~W{-}H Sheu, and K.~C. Sahu.
\newblock Dynamics and stability of a power{-}law film flowing down a slippery
  slope.
\newblock {\em Phys.\ Fluids}, 31:013102, 2019.

\bibitem{Sahu2008}
K.~C. Sahu, A.~Sameen, and R.~Govindarajan.
\newblock The relative roles of divergence and velocity slip in the stability
  of plane channel flow.
\newblock {\em Eur. Phys. J. Appl. Phys.}, 44:101--107, 2008.

\bibitem{Barkley2008}
D.~Barkley, H.~M. Blackburn, and S.~J. Sherwin.
\newblock Direct optimal growth analysis for timesteppers.
\newblock {\em Int. J. Numer. Meth. Fluids}, 57:1435--1458, 2008.

\bibitem{Trefethen2000}
L.~N. Trefethen.
\newblock {\em Spectral methods in Matlab}.
\newblock SIAM, 2000.

\bibitem{Hugues1998}
Sandrine Hugues and Anthony Randriamampianina.
\newblock An improved projection scheme applied to pseudospectral methods for
  the incompressible navier-stokes equations.
\newblock {\em Int. J.\,Num.\ Meth.\ Fluids}, 28:501--521, 1998.

\bibitem{Jimenez2013}
J.~Jimenez.
\newblock How linear is wall{-}bounded turbulence.
\newblock {\em Phys.\ Fluids}, 25:110814, 2013.

\bibitem{Sahu2011}
K.~C. Sahu and O.~K. Matar.
\newblock Three{-}dimensional convective and absolute instabilities in
  pressure{-}driven two{-}layer channel flow.
\newblock {\em International Journal of Multiphase Flow}, 37:987--993, 2011.

\end{thebibliography}

\end{document}